\documentstyle[12pt]{article}
\def\del{\partial}
\def\tr{{\rm tr}\ }
\def\boldalpha{\mbox{\boldmath $\alpha$}}
\def\boldomega{\mbox{\boldmath $\omega$}}
\def\boldA{\mbox{\boldmath $A$}}
\def\boldB{\mbox{\boldmath $B$}}
\def\boldc{\mbox{\boldmath $c$}}

\def\boldF{\mbox{\boldmath $F$}}
\def\boldV{\mbox{\boldmath $V$}}
\def\boldW{\mbox{\boldmath $W$}}
  \makeatletter
        \@addtoreset{equation}{section}
        \def\section{\@startsection {section}{1}{\z@}{3.5ex plus -1ex minus
        -.2ex}{2.3ex plus .2ex}{\large\bf}}
  \makeatother
\begin{document}
\setlength{\baselineskip}{22pt}
\rightline{YAMAGATA-HEP-95-12}
\vspace{1.5truecm}
%
%%%% change the footnote mark into symbol %%%%
\renewcommand{\thefootnote}{\fnsymbol{footnote}}
%%%%%%%%%%%%%%%%%%%%%%%%%%%%%%
%%
\centerline{\large \bf BRST Symmetric Gaugeon Formalism for Yang-Mills Fields%
\footnote{%%%%%%%%%%%%%%%%%%%%%%%
          Preliminary result was reported at 48th Annual Meeting of
          Physical Society of Japan held on March 29 -- April 1, 1993,
          in Sendai.
          }
           }
\vspace{2truecm}
\centerline{Minoru KOSEKI\footnote{%
            Present address: Graduate School of Science and Technology,
            Niigata University, Niigata 950-21, Japan.},
       Masaaki SATO\footnote{%
            Present address: Hitachi Tohoku Software Co., Sendai 980, Japan.}
       and Ryusuke ENDO}
\bigskip
\centerline{\it Department of Physics, Yamagata University, Yamagata 990,
Japan}
\bigskip
\vspace{2truecm}
%
%%%% change back the footnote mark into arabic and reset counter %%%
\renewcommand{\thefootnote}{\arabic{footnote}}
\setcounter{footnote}{0}
%%%%%%%%%%%%%%%%%%%%%%%%%%%%%%%%%%%
%
\centerline{\bf Abstract}
\bigskip
Yokoyama's gaugeon formalism is knwon to admit
$q$-number gauge transformation.
We introduce BRST symmetries into the formalism
for the Yang-Mills gauge field.
Owing to the BRST symmetry, Yokoyama's physical subsidiary conditions
are replaced by a single condition of the Kugo-Ojima type.
Our physical subsidiary condition is
invariant under the $q$-number gauge transformation.
Thus, our physical subspace is gauge invariant.
\thispagestyle{empty}
%
%section1
         \newpage
%section1
%%%%%%%%%%%%%%%%
\section{Introduction}
In the standard formalism of canonically quantized gauge theories \cite{N,KO}
we
%cannot consider the gauge transformation freely.
do not consider the gauge transformation which connects
field operators of different gauges.
There are no such gauge freedom  in the quantum theory since the quantum
theory is defined only after the gauge fixing. In other words,
the  Fock space defined in a particular gauge is not wide enough
%if we want
to realize the quantum gauge freedom.
%
%quite different from those in other gauges.
%Thus, if we want to realize the quantum gauge freedom, we need a wider
%Hilbert space.

Yokoyama's gaugeon formalism \cite{OEM}-\cite{YM2} provides a wider
framework in which we can consider the quantum gauge transformation
among a family of Lorentz covariant linear gauges. In this formalism
a set of extra fields, so called gaugeon fields, is introduced as the quantum
gauge freedom. This theory was first proposed for the quantum
electrodynamics \cite{OEM,YK,txt}
to resolve the problem of gauge parameter renormalization \cite{HY}.
It was also applied later to the Yang-Mills theory \cite{YM1,YM2}.
Owing to the quantum gauge freedom it becomes very easy to check
the gauge parameter independence of the physical $S$-matrix
%becomes manifest
\cite{Smatrix}.
The gauge dependence of the wave-function renormalization constant
was also investigated  %by taking the advantage of
in
this formalism \cite{Zfactor}.

We should ensure that the gaugeon modes do not contribute to the physical
processes. In fact, the gaugeon fields yield negative normed states
that would lead to the negative probability \cite{OEM}.
To remove these unphysical gaugeon modes Yokoyama
imposed a Gupta-Bleuler type subsidiary condition \cite{OEM,YM1,YM2}.
However, this type of condition is not
applicable if interaction exists for the gaugeon fields.
Especially, we cannot use the condition in the background gravitational
field.

Yokoyama's subsidiary condition can be improved if we can introduce
the Becchi-Rouet-Stora-Tyutin (BRST) symmetry \cite{BRST}
for the gaugeon fields.
Izawa has proposed a BRST symmetric Lagrangian for the gaugeon
formalism in the quantum electrodynamics (QED) \cite{Izawa}.
Independently of Izawa's work, we also have presented a BRST
symmetric gaugeon formalism for the QED \cite{KSE}.
Both theories\footnote{
        For the relation between Izawa's theory and ours,
        see Refs.\cite{KSE,RE}.
                       }
include  Faddeev-Popov (FP) ghosts for the gaugeon fields
as well as the usual FP ghosts. As a result, the theories
have larger BRST symmetry and corresponding conserved charges
(BRST charges). Using the BRST charges, we can replace the
Yokoyama's subsidiary condition by a single Kugo-Ojima type condition
\cite{KO},
which is applicable even to the interacting case.

In the present paper, we extend our BRST symmetric gaugeon formalism
for QED to the Yang-Mills gauge theories. We do this by simply
introducing BRST symmetry into the original gaugeon formalisms for the
Yang-Mills fields.
There are two types of gaugeon formalisms for
Yang-Mills fields so far. One of them was proposed by Yokoyama \cite{YM1}.
It has a group vector valued gauge fixing parameter
$\boldalpha=(\alpha^a)$.
The gauge fixing is different from the standard one in the sense that
it breaks not only the local gauge symmetry but also
the rigid gauge symmetry.
The other type of the formalism was
proposed by Yokoyama, Takeda and Monda \cite{YM2}. It
has a (group scalar valued) single gauge fixing parameter $\alpha$.
Thus
the gauge fixing does not violate the rigid gauge symmetry;
though the Lagrangian has nonpolynomial interaction terms.
In the present paper we introduce %the
larger BRST symmetry into
both types of the gaugeon
formalism for the Yang-Mills fields.

The notation and convention used in this paper are the following.
The metric we  use is $g_{\mu\nu}={\rm diag}(+1,-1,-1,-1)$.
The gauge group we consider is a $n$-dimensional compact Lie
group, the generators of which are denoted by $T^a$ ($a=1,2,\dots n$).
Latin letters $a, b, c,\dots$ denote the group vector indices,
while Greek letters $\mu, \nu, \lambda,\dots$ express the space-time
indices which run from 0 to 3. The summation convention is assumed
for both group vector indices and space-time indices. The generators
satisfy
$$
    (T^a)^\dagger = T^a, \qquad   [T^a, T^b] = i f^{abc} T^c.
$$
Here the structure constant $f^{abc}$ is totally antisymmetric
since we assume the normalization for the generators as
$$
          \tr T^a T^b = {1\over2} \delta^{ab}.
$$
%
%section2
         \newpage
%section2
%%%%%%%%%%%%%%%%
\section{Gaugeon formalism with a group vector
valued gauge parameter}
In the formalism we discuss  in this section, the group vector
valued gauge fixing parameter $\mbox{\boldmath $\alpha$}=(\alpha^a)$
is introduced.\footnote{
    We use the group vector notation in this section:
    Letters in boldface denote group vectors. For any two group
    vectors $\boldV=(V^a)$ and $\boldW=(W^a)$,
    we have an inner product
    $$
          \mbox{\boldmath $V$} \mbox{\boldmath $W$}= V^a W^a,
    $$
    and an exterior product
    $$
        \mbox{\boldmath $V$} \times \mbox{\boldmath $W$}
           =  f^{abc} V^b W^c.
    $$
                        }
As a result, Yokoyama's gaugeon fields $Y$ and $Y_*$
are group scalar, while the Nakanishi-Lautrup (Lagrange
multiplier) field $\boldB=(B^a)$ and FP-ghost fields
$\boldc = (c^a)$ and
$\boldc_*=(c^a_*)$ are group vector valued.

\subsection{Yokoyama's theory}
Yokoyama's Lagrangian for the Yang-Mills field
$\boldA_\mu=(A^a_\mu)$ is given by
\begin{eqnarray}
    L_{\rm Y} &=&
        - {1\over4}\boldF^{\mu\nu} \boldF_{\mu\nu}
        - \boldA^{\mu} \nabla_\mu \boldB
        + \del^\mu Y_* \del_\mu Y
        + {\varepsilon \over 2}
              ( Y_* + \boldalpha \boldB)^2
    \nonumber  \\
     & & -i \nabla^\mu \boldc_*
               D_\mu \boldc
        + L_{\rm matter}(\psi, D_\mu\psi),
     \label{L_yokoyama}
     \\
     \boldF_{\mu\nu}
        &=& \del_\mu \boldA_\nu
         - \del_\nu \boldA_\mu
         + g \boldA_\mu \times \boldA_\nu,
    \label{Fmn}
    \\
    D_\mu \boldc
         &=& \del_\mu \boldc
         + g \boldA_\mu \times \boldc,
    \qquad
    D_\mu \psi = (\del_\mu - ig A^a_\mu T^a) \psi,
    \label{Dm}
    \\
    \nabla_\mu \boldV
         &=& \del_\mu \boldV
         + g \boldalpha \del_\mu Y \times \boldV,
         \qquad \quad (\boldV=\boldB, \ \boldc_*)
    \label{nabla}
\end{eqnarray}
where $\boldalpha$ is the group vector valued gauge fixing parameter,
$g$ the coupling constant, $\varepsilon$ a sign factor ($=\pm 1$),
$L_{\rm matter}(\psi, D_\mu\psi)$  the Lagrangian of a matter field
$\psi$ minimally coupled with $\boldA_\mu$,
$\boldF_{\mu\nu}$ the field strength,
$Y$ and $Y_*$  the gaugeon field and its associated field
subject to the Bose-Einstein statistics,
$\boldc$ and $\boldc_*$  are the FP-ghost fields subject
to the Fermi-Dirac statistics,
$D_\mu$ is the covariant derivative, and $\nabla_\mu$
is called the {\it form covariant derivative}.
Since the gauge parameter $\boldalpha$ is group vector valued,
the gauge field propagator is different from the
standard one. In fact, the tree level propagator in the
momentum space is given by
\begin{equation}
    \langle A^a_\mu A^b_\nu \rangle
    \sim
     {\delta^{ab} \over k^2}
                    \left(
                         g_{\mu\nu}
                         - {k_\mu k_\nu \over k^2}
                    \right)
    + \varepsilon \alpha^a \alpha^b {k_\mu k_\nu \over (k^2)^2},
\end{equation}
which does not coincide with the propagator of the standard formalism
unless the Landau gauge ($\boldalpha = 0$) is chosen.

The Lagrangian (\ref{L_yokoyama}) admits $q$-number gauge transformations.
Under the infinitesimal field transformation
\begin{eqnarray}
     \hat{\boldA}_\mu
       &=& \boldA_\mu
       + \tau D_\mu (\boldalpha Y)
       = \boldA_\mu
       + \tau(\boldalpha \del_\mu Y
            + g \boldA_\mu
            \times \boldalpha Y),
    \nonumber
    \\
     \hat{\psi} &=& (1 - ig\tau \alpha^a Y T^a)\psi,
    \nonumber
    \\
     \hat{\boldB}
       &=& \boldB
       + \tau g \boldB
            \times \boldalpha Y,
    \label{Qtransformation}
    \\
     \hat{Y}&=&Y,
    \qquad
      \hat{Y}_* = Y_* - \tau \boldalpha \boldB,
    \nonumber
    \\
     \hat{\boldc} &=& \boldc + \tau g \boldc \times \boldalpha Y,
    \qquad
     \hat{\boldc}_* = \boldc_* + \tau g \boldc_* \times \boldalpha Y
    \nonumber
\end{eqnarray}
with $\tau$ being an infinitesimal parameter (group scalar), the
Lagrangian is {\it form invariant}, that is, it transforms as
\begin{equation}
   L_{\rm Y}(\phi^A; \boldalpha)
   = L_{\rm Y}(\hat{\phi}^A; \hat{\boldalpha}),
  \label{forminvariance}
\end{equation}
where $\phi^A$ stands for any of the fields we are considering
and $\hat{\boldalpha}$ is
defined by
\begin{equation}
    \hat{\boldalpha} = (1 + \tau) \boldalpha .
  \label{a_dilatation}
\end{equation}
Similarly,  under the infinitesimal group vector rotation
\begin{eqnarray}
  \hat{\boldA}_\mu &=& \boldA_\mu + \boldA_\mu \times \boldomega,
  \qquad
  \hat{\psi} = (1 -ig \omega^a T^a )\psi,
  \nonumber
  \\
  \hat{\boldB} &=& \boldB + \boldB \times \boldomega,
  \nonumber \\
  \hat{Y} &=& Y, \qquad \hat{Y}_*=Y_*,
  \label{rotation} \\
  \hat{\boldc} &=& \boldc + \boldc \times \boldomega,
  \qquad
  \hat{\boldc}_* = \boldc_* + \boldc_* \times \boldomega
  \nonumber
\end{eqnarray}
with $\boldomega=(\omega^a)$ being an infinitesimal group vector
parameter,
the Lagrangian transforms as (\ref{forminvariance}) with
$\hat{\boldalpha}$ given by
\begin{equation}
  \hat{\boldalpha} = \boldalpha + \boldalpha \times \boldomega.
 \label{a_rotation}
\end{equation}
The form invariance (\ref{forminvariance}) (under (\ref{Qtransformation})
and (\ref{rotation})) means that $\phi^A$ and $\hat{\phi}^A$ satisfy
the same field equation except for the parameter $\boldalpha$ which
should be replaced by $\hat{\boldalpha}$ for the $\hat{\phi}^A$ field
equation. Thus, we can shift and rotate the gauge parameter $\boldalpha$
by the $q$-number gauge transformations (\ref{Qtransformation}) and
(\ref{rotation}). Note that the sign factor $\varepsilon$
cannot be changed by these transformations.

The Lagrangian (\ref{L_yokoyama}) is invariant under the following
BRST transformation:
\begin{eqnarray}
     \delta_{\rm B} \boldA_\mu &=& D_\mu \boldc ,
     \qquad
     \delta_{\rm B}\psi = -ig c^a T^a \psi,
     \nonumber
     \\
     \delta_{\rm B} \boldc &=& - {g \over 2} \boldc \times \boldc,
     \nonumber
     \\
     \delta_{\rm B} \boldc_* &=& i \boldB,
     \qquad
     \delta_{\rm B} \boldB = 0 ,
     \label{BRSTy}
     \\
     \delta_{\rm B} Y  &=& \delta_{\rm B} Y_* = 0,
     \nonumber
\end{eqnarray}
which obviously satisfies the nilpotency,
$
      {\delta_{\rm B}}^2  = 0.
$
Corresponding to this  invariance, there exists a
Noether current $J_{\rm B}^\mu$ satisfying the conservation law
\begin{equation}
              \del_\mu J_{\rm B}^\mu=0.
\end{equation}
Thus we can define the BRST charge by
\begin{equation}
   Q_{\rm B} = \int d^3x J_{\rm B}^0,
\end{equation}
which satisfies the nilpotency ${Q_{\rm B}}^2=0$.

%We should ensure that the unphysical modes of gauge fields and gaugeon
%fields do not contribute to the physical processes.
To remove the unphysical modes and  define physical states,
Yokoyama imposed two kinds of subsidiary conditions:
\begin{equation}
          Q_{\rm B} \left|{\rm phys} \right> =0,
\end{equation}
\begin{equation}
          (Y_* + \boldalpha \boldB)^{(+)}
                        \left| {\rm phys} \right> = 0.
\end{equation}
As shown by Kugo and Ojima \cite{KO}, the first condition removes the
unphysical gauge field modes from the total Fock space.
The nilpotency and conserving property of $Q_{\rm B}$ is essential
in proving that this condition works well.
The second condition is a Gupta-Bleuler type condition \cite{N};
the superscript $(+)$ denotes the positive frequency part.
It removes the unphysical gaugeon modes. In this context, it is
important that the combination
\begin{equation}
     \Lambda = Y_* + \boldalpha \boldB
\end{equation}
satisfies the free field equation
\begin{equation}
     \Box \Lambda = 0 .
\end{equation}
Owing to the free equation,
the decomposition of $\Lambda$ into the positive and negative
frequency parts is well-defined.
However, once we consider the gravitational interaction, the decomposition
of $\Lambda$ into $\Lambda^{(\pm)}$ is no longer well-defined.
This is the limitation of the Gupta-Bleuler type subsidiary condition;
the
Kugo-Ojima type condition based on the conserved charge has no
limitation of this kind.

%%%%%%%%%%%%%%%%%%%%%%%%%%%%%%%%%%%
\subsection{BRST symmetric theory}
As a BRST symmetric version of (\ref{L_yokoyama}) we propose the
following Lagrangian:
\begin{eqnarray}
    L &=&
        - {1\over4}\boldF^{\mu\nu} \boldF_{\mu\nu}
        - \boldA^{\mu} \nabla_\mu \boldB
        + \del^\mu Y_* \del_\mu Y
        + {\varepsilon \over 2}
              ( Y_* + \boldalpha \boldB)^2
    \nonumber  \\
     & & -i \nabla^\mu \boldc_*
               D_\mu \boldc
        -i \del^\mu K_* \del_\mu K
        + L_{\rm matter}(\psi, D_\mu\psi),
  \label{L_ym1}
\end{eqnarray}
where group scalars
$K$ and $K_*$, subject to the Fermi-Dirac statistics, are
FP-ghost fields for the gaugeon fields $Y$ and $Y_*$.

By introducing $K$ and $K_*$ we are able to extend the BRST transformation
so that the gaugeon fields are also transformed. We consider the
following larger BRST transformation:
\begin{eqnarray}
     \delta_{\rm B} \boldA_\mu &=& D_\mu \boldc ,
       \qquad
     \delta_{\rm B}\psi = - ig c^a T^a \psi,
       \nonumber \\
     \delta_{\rm B} \boldc &=& - {g \over 2} \boldc \times \boldc,
       \nonumber  \\
     \delta_{\rm B} \boldc_* &=& i \boldB,
       \qquad
     \delta_{\rm B} \boldB = 0 ,
       \label{QBym1} \\
%       \nonumber \\
     \delta_{\rm B} Y &=& K,
       \qquad
     \delta_{\rm B}K =0,
       \nonumber \\
     \delta_{\rm B} K_*  &=& -i Y_*,
       \qquad
     \delta_{\rm B} Y_* =0,
       \nonumber
\end{eqnarray}
which satisfies ${\delta_{\rm B}}^2=0$. Because of the nilpotency, the
invariance under
this transformation can be easily seen if we rewrite the Lagrangian as
\begin{eqnarray}
  L &=& - {1 \over 4} \boldF^{\mu\nu} \boldF_{\mu\nu}
        + L_{\rm matter}(\psi, D_\mu \psi)
      \nonumber \\
    & & - i \delta_{\rm B} \left[
          \boldc_* \left(
                   \nabla^\mu \boldA_\mu
                    - {\varepsilon \boldalpha \over 2}
                      (Y_* + \boldalpha \boldB)
                   \right)
          + K_*    \left(
                   \Box Y - {\varepsilon \over 2}
                      (Y_* + \boldalpha \boldB)
                   \right)
                            \right].
\end{eqnarray}
The  BRST current is now given by
\begin{equation}
  J_{\rm B}^\mu = - \boldF^{\mu\nu} D_\nu \boldc
              - i \ {g \over 2} \ \nabla^\mu \boldc_* (\boldc \times \boldc)
              - (D^\mu \boldc) \boldB
              - Y_* D^\mu K,
\end{equation}
which yields the conserved BRST charge $Q_{\rm B}=\int d^3x J_{\rm B}^0$.

As for the $q$-number gauge transformation, we now consider the
following field transformation:
\begin{eqnarray}
     \hat{\boldA}_\mu
       &=& \boldA_\mu
       + \tau D_\mu (\boldalpha Y)
    \qquad
     \hat\psi = (1-ig \tau \alpha^a T^a Y )\psi,
    \nonumber
    \\
     \hat{\boldB}
       &=& \boldB
       + \tau g \boldB
            \times \boldalpha Y
       - i \tau g \boldc_* \times \boldalpha K,
    \nonumber
    \\
     \hat{Y}&=&Y,
    \qquad
     \hat{Y}_*
       = Y_* - \tau \boldalpha \boldB,
    \label{QGTym1} \\
    %\nonumber \\
     \hat{\boldc} &=& \boldc + \tau g \boldc \times \boldalpha Y
                     + \tau \boldalpha K,
     \qquad
     \hat{\boldc}_* = \boldc_* + \tau g \boldc_* \times \boldalpha Y
    \nonumber \\
     \hat{K} &=& K,
    \qquad
     \hat{K}_* = K_* - \tau \boldalpha \boldc.
    \nonumber
\end{eqnarray}
Under this transformation [ and the rotation (\ref{rotation})]
the Lagrangian (\ref{L_ym1}) is again form invariant:
\begin{equation}
   L(\phi^A; \boldalpha)
   = L(\hat{\phi}^A; \hat{\boldalpha})
\end{equation}
with $\hat{\boldalpha}$ given by (\ref{a_dilatation})
[or by (\ref{a_rotation})].
Thus the theories with different gauge fixing parameters $\boldalpha$
are included in one theory described by the Lagrangian
(\ref{L_ym1}).

The physical subsidiary condition becomes now simpler. We impose
a single  condition,
\begin{equation}
  Q_{\rm B} \left| {\rm phys} \right> = 0.
\end{equation}
Since our BRST operator acts on the
gaugeon fields as well as usual gauge
fields, the condition removes all the unphysical modes.
(For example, as seen from (\ref{QBym1}),
$Y$, $Y_*$, $K$ and $K_*$ form a
BRST quartet \cite{KO}, which is known to appear only as zero-normed states
in the physical subspace.) Thus we are able to avoid the
Gupta-Bleuler type subsidiary condition. Consequently,
our physical condition works well
even in the background gravitational filed.

%section3
         \newpage
%section3
%%%%%%%%%%%%%%%%
\section{Gaugeon formalism with a single gauge parameter}
In the present section we consider the gaugeon formalism in which
the gauge fixing parameter is a group scalar.
In this sense the theory is more similar with the standard formalism
\cite{KO} than the theory discussed in the last section.
And
gaugeon fields have also group vector indices ($Y^a$ and
$Y_*^a$).
We use the matrix notation
for the group vector in this section.
For any group vector $\boldV=(V^a)$, we define%
\footnote{
      In this notation, a commutator corresponds to the exterior product:
      $
         -i [V, W] =  (\boldV \times \boldW)^a T^a .
      $
      }
$$
          V = V^a T^a.
$$

\subsection{Yokoyama-Takeda-Monda theory}
The Lagrangian of Yokoyama-Takeda-Monda theory \cite{YM2} is given by
\begin{eqnarray}
    L_{\rm YTM} &=& 2 \tr \left\{
          - {1\over4} F^{\mu\nu} F_{\mu\nu}
          + (A^{\mu} - F^\mu) \nabla_\mu B
                          \right\}
       \nonumber \\
      &+&  2 \tr \left\{
          \del^\mu Y_* \del_\mu Y
          + {\varepsilon \over 2} {Y_*}^2
           -i \nabla^\mu c_* D_\mu c
                  \right\}
        + L_{\rm matter}(\psi, D_\mu\psi),
     \label{L_YTM}
\end{eqnarray}
with
\begin{eqnarray}
     F_{\mu\nu}
        &=& \del_\mu A_\nu
         - \del_\nu A_\mu
         - i g [A_\mu, A_\nu],
    \label{Fmn'}
    \\
    D_\mu V
         &=& \del_\mu V
         - i g [A_\mu, V],
    \label{Dm'}
    \\
    \nabla_\mu V
         &=& \del_\mu V
         - i g \alpha [F_\mu, V].
    \label{nabla'}
    \\
    - i g \alpha F_\mu &=& S^{-1}(\alpha)\ \del_\mu S(\alpha),
    \label{Fm}
    \\
    S(\alpha) &=& \exp (-ig\alpha Y),
\end{eqnarray}
where  $\alpha$ is the group scalar valued gauge fixing parameter,
$\varepsilon$ again a sign factor, and $Y=Y^a T^a$ and $Y_*=Y_*^a T^a$
are the Lie algebra valued gaugeon fields.
The tree level propagator of gauge fields is given by
\begin{equation}
    \langle A^a_\mu A^b_\nu \rangle
    \sim
     {\delta^{ab} \over k^2}
                    \left(
                         g_{\mu\nu}
                         + (\varepsilon \alpha^2 - 1)
                         {k_\mu k_\nu \over k^2}
                    \right),
\end{equation}
which coincides with the propagator of the standard formalism though
the nonperturbative propagator differs from the standard one.
Note that $S(\alpha)$ has its value in the group
and consequently $F_\mu (=F_\mu^a T^a)$
lives in the Lie algebra.
As seen from (\ref{Fm}), $F_\mu$ is a nonpolynomial function of $Y$.
The renormalizability of this theory with such nonpolynomial interactions
is  discussed in Ref.\cite{YM2}.

The Lagrangian admits the
$q$-number gauge transformation defined by
\begin{eqnarray}
     \hat{A}_\mu
       &=& S^{-1}(\tau)\ A_\mu S(\tau)
         + {i \over g}  S^{-1}(\tau)\  \del_\mu S(\tau),
    \nonumber
    \\
     \hat{\psi} &=& S^{-1}(\tau)\ \psi,
    \nonumber
    \\
     \hat{V}
       &=& S^{-1}(\tau) \ V S(\tau), \qquad \mbox{($V=B$, $c$, $c_*$)}
    \label{QGTytm}
    \\
     \hat{W} &=& W, \qquad \qquad \mbox{($W=Y$, $Y_*$)}
    \nonumber
\end{eqnarray}
with $\tau$ being a finite parameter.
Under this field transformation, the Lagrangian is form invariant:
\begin{equation}
   L_{\rm YTM}(\phi^A; \alpha)
   = L_{\rm YTM}(\hat{\phi}^A; \hat{\alpha}),
  \label{forminvarianceYTM}
\end{equation}
where $\phi^A$ stands for any of the fields and $\hat{\alpha}$ is
defined by
\begin{equation}
    \hat{\alpha} = \alpha +  \tau.
  \label{alphahat}
\end{equation}

%%%%%%%%%%%%%%%%%%%%%%%%%%
The Lagrangian (\ref{L_YTM}) is invariant under the following
BRST transformation:
\begin{eqnarray}
     \delta_{\rm B} A_\mu &=& D_\mu c ,
     \qquad
     \delta_{\rm B}\psi = -igc\psi,
     \nonumber \\
     \delta_{\rm B} B &=& \delta_{\rm B} Y = \delta_{\rm B}Y_*=0 ,
     \label{BRSTytm} \\
     \delta_{\rm B} c &=& i g c^2,
     \qquad
     \delta_{\rm B} c_*  = i B,
     \nonumber
\end{eqnarray}
which obviously satisfies the nilpotency,
$
      {\delta_{\rm B}}^2  = 0.
$
Corresponding to this symmetry, we have a conserved  BRST charge
$Q_{\rm B}$.

To remove the unphysical modes and define physical states,
Yokoyama, Takeda and Monda imposed the following conditions:
\begin{eqnarray}
      & &Q_{\rm B} \left|{\rm phys} \right> =0,
      \label{Q_bytm} \\
      & &Y_*^{(+)} \left| {\rm phys} \right> = 0.
      \label{G-Bytm}
\end{eqnarray}
The first condition removes the unphysical modes of gauge field while
the second eliminates  unphysical gaugeon modes.
It is essential in the  second condition
that the field $Y_*$ satisfies the free field equation,
\begin{equation}
     \Box Y_* = 0 .
\end{equation}
Owing to the free equation, the decomposition of $Y_*$ into the positive
and negative frequency parts $Y_*^{(\pm)}$ is well-defined.
However, once we consider the gravitational interaction, the Gupta-Bleuler
type condition (\ref{G-Bytm}) no longer works well.

%%%%%%%%%%%%%%%%%%%%%%%%%%%%%%%%%%%%
\subsection{BRST symmetric theory}
As a BRST symmetric version of (\ref{L_YTM}) we present a
Lagrangian given by
\begin{eqnarray}
    L &=& 2 \tr \left\{
          - {1\over4} F^{\mu\nu} F_{\mu\nu}
          + (A^{\mu} - F^\mu) \nabla_\mu B
          + \del^\mu Y_* \del_\mu Y
          + {\varepsilon \over 2} {Y_*}^2
                          \right\}
       \nonumber \\
      &+ &  2 \tr \left\{
           -i \nabla^\mu c_* D_\mu c
           -i \del^\mu K_* \del_\mu K
                  \right\}
        + L_{\rm matter}(\psi, D_\mu\psi),
     \label{Lym2}
\end{eqnarray}
where
$K=K^a T^a$ and $K_*=K_*^a T^a$ have been introduced as Lie algebra valued
FP-ghost fields for the gaugeon fields $Y$ and $Y_*$.

We may consider the $q$-number transformation defined by
\begin{eqnarray}
     \hat{A}_\mu
       &=& S^{-1}(\tau)\ A_\mu S(\tau)
         + {i \over g} S^{-1}(\tau)\ \del_\mu S(\tau),
    \nonumber
    \\
     \hat\psi &=& S^{-1}(\tau)\ \psi,
    \nonumber
    \\
     \hat{V}
       &=& S^{-1}(\tau)\ V S(\tau),
    \label{QGTym2}
    \\
     \hat{W} &=& W,
    \nonumber
\end{eqnarray}
where, and in the following, $V$ stands for $B$, $c$, and $c_*$ and
$W$ denotes $Y$, $Y_*$, $K$, and $K_*$.
The Lagrangian is form invariant under this transformation:
\begin{equation}
    L(\hat \phi^A; \hat\alpha) = L(\phi^A; \alpha)
    \label{forminvYM2}
\end{equation}
with $\hat\alpha$ being $\hat{\alpha}=\alpha + \tau$.
To check the form invariance  (\ref{forminvYM2})
we have used the identities,
\begin{eqnarray}
 & & \hat{A}_\mu - \hat\alpha \hat F_\mu =
    S^{-1}(\tau)\ (A_\mu - \alpha F_\mu) S(\tau),
    \nonumber \\
 & & \hat F_{\mu\nu} = S^{-1}(\tau)\ F_{\mu\nu} S(\tau),
   \nonumber \\
 & & \widehat{\nabla_\mu V} = S^{-1}(\tau)\ \nabla_\mu V S(\tau),
   \label{QGTym2'} \\
 & & \widehat{D_\mu V} = S^{-1}(\tau)\ D_\mu V S(\tau).
   \nonumber
\end{eqnarray}

The BRST transformation we propose here is
\begin{eqnarray}
     \delta_{\rm B} A_\mu &=& D_\mu (c + \alpha {\cal K}) ,
       \nonumber \\
     \delta_{\rm B}\psi &=& -ig( c + \alpha {\cal K})\psi,
       \nonumber \\
     \delta_{\rm B}B  &=& i g \alpha [{\cal K}, B] ,
       \nonumber \\
     \delta_{\rm B} c &=& ig \left\{%%%%%%%
                        {1\over2}c + \alpha {\cal K}, c
                             \right\}, %%%%%%
       \nonumber  \\
     \delta_{\rm B} c_* &=& -i B + ig\alpha \{{\cal K}, c_* \},
       \nonumber \\
     \delta_{\rm B} Y  &=& K,
       \label{QBym2} \\
     \delta_{\rm B} Y_* &=& 0,
       \nonumber  \\
     \delta_{\rm B}K &=& 0,
       \nonumber \\
     \delta_{\rm B} K_*  &=& -i Y_*,
       \nonumber
\end{eqnarray}
where ${\cal K}$ is defined by
\begin{eqnarray}
   {\cal K} &=& K^a F^a,
   \\
   -ig\alpha F^a &=& S^{-1}(\alpha)\ {\del \over \del Y^a} S(\alpha).
\end{eqnarray}
By using the identities
\begin{eqnarray}
  & & \delta_{\rm B} F_\mu = \nabla_\mu {\cal K},
   \\
  & & \delta_{\rm B} {\cal K} = ig\alpha {\cal K}^2,
\end{eqnarray}
we can easily check the nilpotency of our BRST transformation (\ref{QBym2}).
Furthermore we can show the BRST invariance of the Lagrangian since
we may rewrite the Lagrangian as
\begin{eqnarray}
   L &=& 2\tr \left\{
        - {1\over4} F^{\mu\nu} F_{\mu\nu}
              \right\}
        + L_{\rm matter}(\psi, D_\mu\psi)
     \nonumber
     \\
     & & -i \delta_{\rm B} \left[
                                 2\tr \left\{
              c_* \del^\mu(A_\mu - \alpha F_\mu)
              - \del^\mu K_* \del_\mu Y
              - {\varepsilon \over 2}K_* Y_*
                                  \right\}
                           \right].
\end{eqnarray}
We have thus a conserved and nilpotent
BRST charge $Q_{\rm B}$ corresponding to the symmetry under (\ref{QBym2}).
Using the BRST charge we impose the physical subsidiary condition as
\begin{equation}
       Q_{\rm B} \left| {\rm phys} \right> = 0,
\end{equation}
by which we replace the two subsidiary conditions (\ref{Q_bytm}) and
(\ref{G-Bytm}) of Yokoyama, Takeda and Monda.
In particular, we do not need any Gupta-Bleuler type subsidiary condition.
Consequently, our theory is applicable even in the background gravitational
field.
%
%section4
         \newpage
%section4
%%%%%%%%%%%%%%%%
\section{Summary and remarks}
We have presented two kinds of gaugeon formalisms for Yang-Mills fields with
larger BRST symmetries. One is an extension of Yokoyama's theory \cite{YM1}
in which a group vector valued parameter is used in the gauge fixing
term. The other is an extension of the theory by
Yokoyama, Takeda and
Monda \cite{YM2} which has a group scalar valued gauge fixing parameter.
By using  BRST charges corresponding to the larger BRST symmetries,
we have been able to replace the Yokoyama's physical subsidiary conditions by
a single Kugo-Ojima type condition in each case. As a result, the
formalism becomes applicable to the
case of the background gravitational field.

We emphasize that in both cases (of sections 2 and 3)
our physical condition is invariant under the
$q$-number gauge transformation. As seen from (\ref{QBym1}) and
(\ref{QGTym1}), % in section 2,
or %as seen
from (\ref{QBym2}) and
(\ref{QGTym2}), the BRST transformation and the $q$-number gauge
transformation commute with each other. This fact leads us to
\begin{equation}
    \hat{Q}_{\rm B} = Q_{\rm B},
\end{equation}
that is, the BRST charge is invariant under the $q$-number
gauge transformation.
Consequently, our physical
subsidiary conditions, and thus, our physical subspace are
gauge invariant.
In the case of quantum electrodynamics, this kind of structure of
the physical subspace plays an essential role in the proof of the
gauge parameter independence of the physical $S$-matrix \cite{REtalk}.

\bigskip
\bigskip
%\noindent
\leftline{\bf Note added}
After completing this paper, we
were informed of the work by M. Abe
(``The Symmetries of the Gauge-Covariant Canonical Formalism
of Non-Abelian Gauge Theories", Master Thesis, Kyoto University, 1985)
in which he already proposed and studied
the BRST-symmetrized Yokoyama-Takeda-Monda theory.
His Lagrangian and BRST symmetry are the same as ours
discussed in the section 3.
%
%%%
         \newpage
%%%%%%%%%%%%%%%%%%%%
%References
%%%%%%%%%%%%%%%%%%%%

%
%
\end{document}